# Integrating of continuous graphene with periodic ferroelectric domains for adaptive terahertz photodetector

Lin Lin[1,2,3], Junxiong Guo[1,2,5]*, Shangdong Li[1], Tianxun Gong[1], Juan Xia[3], Yang Zhang[4], Wenjing Jie[6], Wen Huang[1]* and Xiaosheng Zhang[1]*

1. School of Electronic Science and Engineering (National Exemplary School of Microelectronics), University of Electronic Science and Technology of China, Chengdu 610054, China

2. School of Electronic Information and Electrical Engineering, Chengdu University,

3. Institute of Fundamental and Frontier Sciences, University of Electronic Science and Technology of China, Chengdu 610054, China

Chengdu 610106, China

4. Tianjin Key Laboratory of Micro-Scale Optical Information Science and Technology, Institute of Modern Optics, Nankai University, Tianjin 300071, China

5. Chengdu Research Institute of UESTC, Chengdu, 610207, China

6. College of Chemistry and Materials Science, Sichuan Normal University, Chengdu, 610068, China

* Corresponding authors

E-mail addresses: guojunxiong@cdu.edu.cn (J.X. Guo); uestchw@uestc.edu.cn (W. Huang) and xszhang@uestc.edu.cn (X.S. Zhang)

## Abstract

Graphene plasmons hold immense potential for terahertz (THz) detector application due to their fascinating interactions between radiation and matter. However, it has remained challenging to excite and manipulate graphene plasmons within continuous graphene that is free of patterning technique. Here, we report an adaptive wavelength-sensitive terahertz detector consisting of continuous graphene integrated onto a ferroelectric thin film with periodic polarization domains. This designed device is capable of absorbing THz waves with zero input bias voltage because of highly confinement of surface plasmons within the interface between graphene and ferroelectrics. By reconfiguring an interweaving squared ferroelectric domain array with alternating upward and downward polarizations, our devices theoretically own an ultrahigh responsivity of 17.56 A $W^{-1}$ and a specific detectivity of $1.11\times10^{11}$ Jones at room temperature. We also demonstrate that the photodetectors make possible for spectrum reconstruction application of portable spectrometer at a broad operation band of 4.97 to 7.85 THz with resolution up to 0.02 THz combining the mathematical algorithms.

Terahertz (THz) detectors are essential for various applications, including communication, astronomical observations, imaging, and macromolecular detection[1-6]. However, these promising applications are impeded from a lack of high-performance THz detectors which are crucial to capture weak THz signals and convert them to electrical ones for following processing. To overcome these restrictions and meet the demands of portable applications, a high-performance and adaptive THz detector with a wide operating frequency range and compact size is highly expected. Recent developments of two-dimensional materials have shown their promising potentials for photodetectors with features of electrically tunable photoresponse and high performance due to the strong light-matter interactions within atomic-layer thickness[7-11]. Graphene, in particular, exhibits an ultrabroad photoresponse across visible to terahertz frequencies, excellent tunability of induced surface plasmon polaritons (SPPs), and outstanding compatibility with the standard technique of complementary metal–oxide–semiconductor[12-16]. To date, many prototype THz detectors based on graphene surface plasmon resonance (SPR) using patterned graphene or applying integrated local electrode gate have already been reported[13, 17]. However, most of these devices are confronted with the narrow operating frequency, and inevitably suffer from poor responsivity induced by incomplete graphene SPR caused by edge disorder[17]. For these reasons, securing an alternative route for efficient SPPs excitation and modulation within continuous graphene sheet has become a widely pursued goal.

Ferroelectric with naturally ultrahigh local electric field of ~$10^9$ V $m^{-1}$ in nanoscale[18-20], combined with the development of ferroelectric polarization switching technologies including ferroelastic switching[21], mechanical writing[22] and water type-printing[23], provide an ideal platform to effectively induce graphene SPR and study graphene plasmons excitation and confinement.

In this work, we propose a terahertz detector based on graphene plasmons excited by reconfigurable ferroelectric domains array, which is free of complex electrode structure and patterned graphene. Based on simulation results, a highly strong SPR effect induced by periodic interweaving squared domains array could be obtained in the proposed detector. This effect could facilitate the absorption of the graphene in the terahertz region and thus enhance photoelectric conversion capabilities of this device. The simulations show that we achieve an ultrahigh responsivity 17.56 A $W^{-1}$ and a detectivity $1.11\times10^{11}$ Jones (1 Jones = 1 cm $Hz^{1/2}$ $W^{-1}$) under 6.30 THz waves at room temperature. This device also features an adaptive absorption peak from 4.97 to 7.85 THz by varying the ferroelectric-controlled graphene chemical potential and scaling the ferroelectric domain size. We also show that the conceptual device can be used for spectrum reconstruction by introducing mathematical algorithms into the portable spectrometer.

As shown in Fig. 1(a), Au electrode, graphene, ferroelectric material, bottom electrode and substrate constitute the graphene plasmonic terahertz detector from top to bottom. To fabricate the photodetector, graphene can be transformed onto the ferroelectric film by mechanical exfoliation or wet method[24, 25]. Epitaxial grown the ferroelectric film ($BiFeO_3$) on a structure including a substrate ($SrTiO_3$) and an electrode ($La_{1/3}Sr_{2/3}MnO_3$) is available as previously reported[22]. The ferroelectric film can be switched to 90°-assigned domains (upward domains) or −90°-assigned domains (downward domains) periodically, which can be realized conveniently by using piezoresponse force microscopy (PFM) method[26]. Also, using photoetch and water writing is another method to realize switching ferroelectric domains periodical with ultralow cost [23]. Through designing size and shape of ferroelectric domains to match the frequency of incident waves, the graphene is able to induce SPPs generally located at the boundary of different domains[27]. In this device whose electrical and optical characteristics are simulated by using finite element method, each ferroelectric domain is square with the same side length (SL) and every two adjacent domains are inversely polarized. Affected by the ferroelectric domains, the position of graphene Dirac points is presented as a gray curved surface in Fig. 1(b) and energy bands of graphene on different ferroelectric domains are also displayed corresponding to different Fermi level. The raised surface corresponds to graphene upon an upward ferroelectric domain and the sunken one corresponds to graphene upon a downward ferroelectric domain. In addition, the mechanism of carriers tunneling between graphene on two same polarized ferroelectric domains is demonstrated in Fig. 1(b).

The fermi level of graphene can be controlled by polarized ferroelectric domains, since the remanent polarization ($P_r$) of polarized ferroelectric domains influences the carrier distribution of graphene and creates an electrostatic potential difference $\phi$. The effective doping concentration of graphene follows the formula as,

$$P_r / C_d = E_f / e + \phi \quad (1)$$

where $E_f/e$ is determined by the quantum capacitance of graphene and $\phi$ depends on geometrical capacitance ($C_d$). For example, with the variation of capacitance ranging from 2 to 4 μC cm$^{-2}$ and the remanent polarization ranging from 0.2 to 1.4 μF cm$^{-2}$, the fermi level of graphene ranges from 0.02 to 0.7 eV, supposing the electrostatic potential difference can be ignored as shown in Fig. 1(c). The chemical potential of graphene ($\mu_c$), which is a key variable to determine the optical conductivity of graphene, is almost equal to the fermi level of graphene ($E_f$) at room temperature and both of them are relevant to the carrier concentration of graphene $N_0$ expressing by Eqn. (2),

$$\mu_c = E_f = \hbar v_f \left( \pi N_0 \right)^{1/2} \quad (2)$$

where $v_f$ is the Fermi velocity of $1.1\times10^6$ m s$^{-1}$.

The optical conductivity of graphene is vital and determine the plasmonic and optics properties of graphene, since it contains all the relevant information of the electromagnetic interactions between graphene and external stimulus [28-31]. The conductivity of graphene ($\sigma_g$) is divided into two parts, the intraband conductivity ($\sigma_{intra}$) presenting the transitions within the valence band or conduction band and interband conductivity ($\sigma_{inter}$) presenting the transitions from the valence band to the conduction band vertically [29, 32], as expressing by $\sigma_g = \sigma_{intra} + \sigma_{inter}$.

Here, the intraband and interband conductivities of graphene are calculated following the Falkovsky formula, as shown in Eqns. (3) and (4) [30, 31],

$$\sigma_{\text{intra}} = \frac{2\mathrm{i}e^2 k_{\mathrm{B}} T}{\pi\hbar^2\left(\omega + \mathrm{i}\tau^{-1}\right)} \ln\left[2\cosh\left(\frac{\mu_{\mathrm{c}}}{2k_{\mathrm{B}}T}\right)\right] \tag{3}$$

$$\sigma_{\text{inter}} = \frac{e^2}{4\hbar}\left[G\left(\frac{\hbar\omega}{2}\right) + \mathrm{i}\frac{4\hbar\omega}{\pi}\int_0^{\infty} \mathrm{d}\varepsilon \frac{G(\varepsilon) - G\left(\frac{\hbar\omega}{2}\right)}{(\hbar\omega)^2 - 4\varepsilon^2}\right] \tag{4}$$

where $\tau$ is the relaxation time, $\omega$ is the radian frequency, $T$ is the temperature, $e$ the elementary charge and $\mu_{\mathrm{c}}$ the chemical potential of graphene. In Eqn.(3), $\tau$ can be expressed as : $\frac{1}{2\Gamma} = \tau = \frac{1}{2\pi}\frac{\mu E_{\mathrm{f}}}{e v_{\mathrm{f}}^2}$ , where $\Gamma$ is scattering rate and $\mu$ is the electron mobility[16, 33, 34]. $G(\varepsilon)$ in Eqn. (4) is another function to present the difference of the Fermi functions, as shown in Eqn. (5).

$$G(\varepsilon) = \frac{\sinh\left(\frac{\varepsilon}{k_{\mathrm{B}}T}\right)}{\cosh\left(\frac{\mu_{\mathrm{c}}}{k_{\mathrm{B}}T}\right) + \cosh\left(\frac{\varepsilon}{k_{\mathrm{B}}T}\right)} \tag{5}$$

In general, the real part of optical conductivity ($\sigma_{\mathrm{gr}}$) is the key ingredient in graphene electrical properties and the imaginary part ($\sigma_{\mathrm{gi}}$) determines which mode of electromagnetic wave can excite the SPPs in graphene. Graphene is able to induce the SPPs under transverse-magnetic waves when $\sigma_{\mathrm{gi}}$ is positive and the SPPs under transverse-electric waves when $\sigma_{\mathrm{gi}}$ is negative [35, 36]. Figs. 1 (d) and (e) show these two parts of optical conductivity as a function between the chemical potential and the frequency of incident waves. The graphene upon downward domains and upward domains behaves as p-doped graphene and near-intrinsic graphene respectively, which is illustrated by previous reports [37, 38].

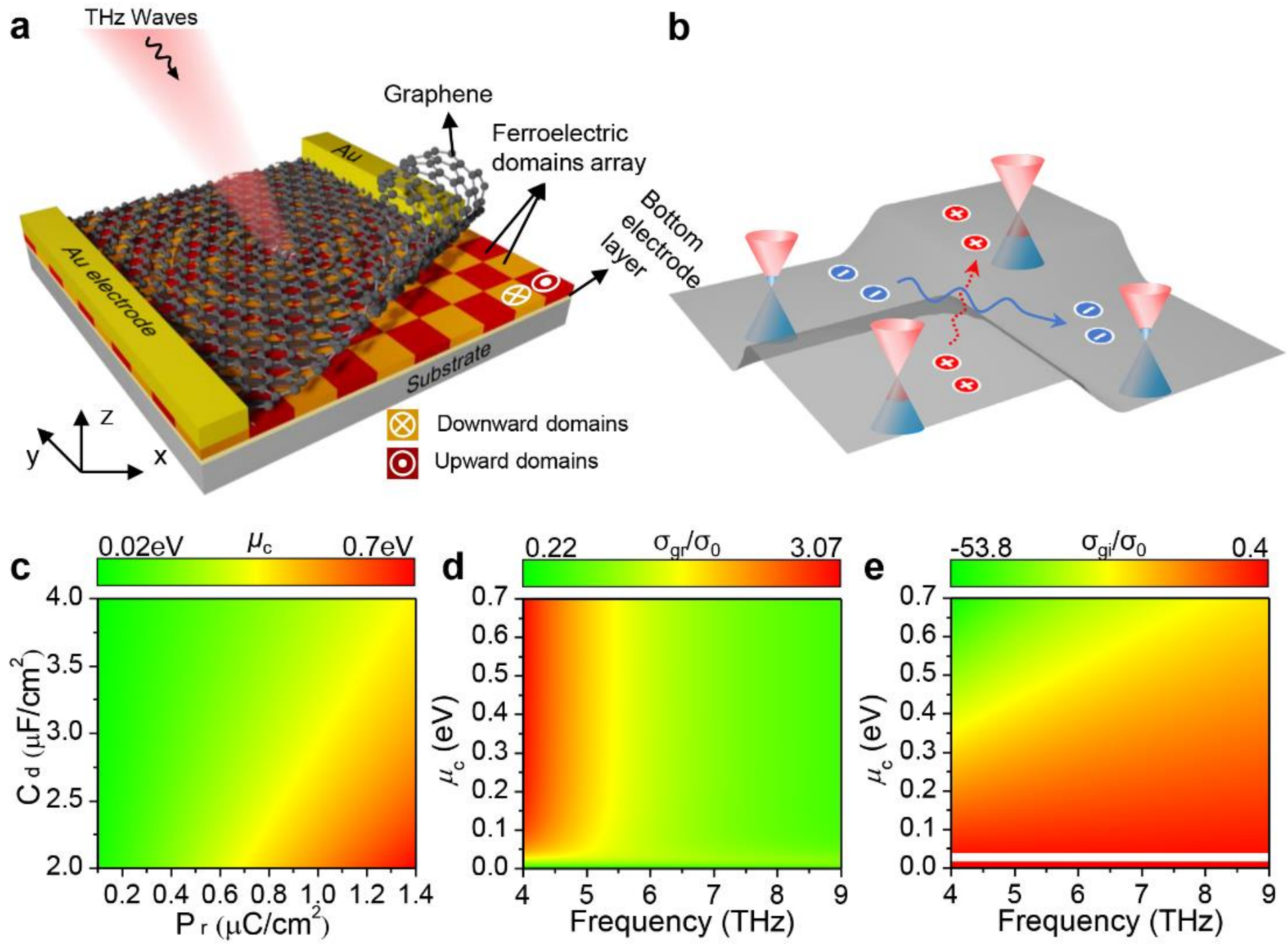

**Fig. 1.** Graphene/ferroelectric-based plasmonic terahertz detector. (a) Schematic of terahertz detector based on integrating of graphene with ferroelectric. (b) Position of Dirac point presented by the grey surface and schematic diagram of carriers tunneling. The sunken and raised planes correspond to graphene upon downward and upward ferroelectric domains respectively. (c) Chemical potential ($\mu_c$) of graphene as a function of the remanent polarization ($P_r$) of the ferroelectric domain and capacitance ($C_d$). The normalized graphene conductivity ($\sigma_g/\sigma_0$), including the real part ($\sigma_{gr}/\sigma_0$) (d) and the imaginary part ($\sigma_{gi}/\sigma_0$) (e), as a function of the chemical potential ($\mu_c$) of graphene and the frequency of incident wave (from 4 to 9 THz), by calculating the Falkovsky formula. The white line in (e) presents the value of $\sigma_{gi}/\sigma_0$ as 0.

The optical properties of graphene-based terahertz detector are simulated through finite element method. Fig. 2(a) shows the simulation unit and its pivotal parameter SL, which is the only one to determine the size of square ferroelectric domains. In this situation, we set the boundary condition both x-axis and y-axis as periodic. The absorption ($A_g$) of the proposed device is quite relevant to the reflectance ($R_g$) and the transmittance ($T_g$), following the formula as $A_g+R_g+T_g=1$. Here, supposing the SL is 1 μm and the chemical potential of graphene upon downward and upward polarized ferroelectric domains are 400 meV and 1 meV, respectively. An electric field with ultrahigh intensity exists at the edge of the graphene upon the upward polarized ferroelectric domains and disappears at graphene located at downward polarized ferroelectric domains, as shown in Fig. 2(b). It indicates the graphene SPPs are confined at the edge of near-intrinsic graphene, and this phenomenon can enormously enhance the absorption of THz waves and the generation of charge carriers. In detail, the absorption of 6.30 THz waves is up to 31%, which is 20 times more than the absorption at 4.5 THz.

As shown in Fig. 2(c), assuming the SL is 1 μm, with the chemical potential of graphene ranging from 0.2 to 0.7 eV influenced by polarized ferroelectric domains, the absorption peak of proposed device shifts from 5.05 to 7.80 THz. Fig. 2(d) reveals the variation of absorption peak shifts from 7.85 to 4.97 THz depending on the SL ranging from 0.7 to 1.5 μm when the chemical potential is set as 0.4 eV. The shape of polarized ferroelectric domains and the remanent polarization can be easily changed through applying different direction and intensity of electric field by PFM, which makes our device reconfigurable.

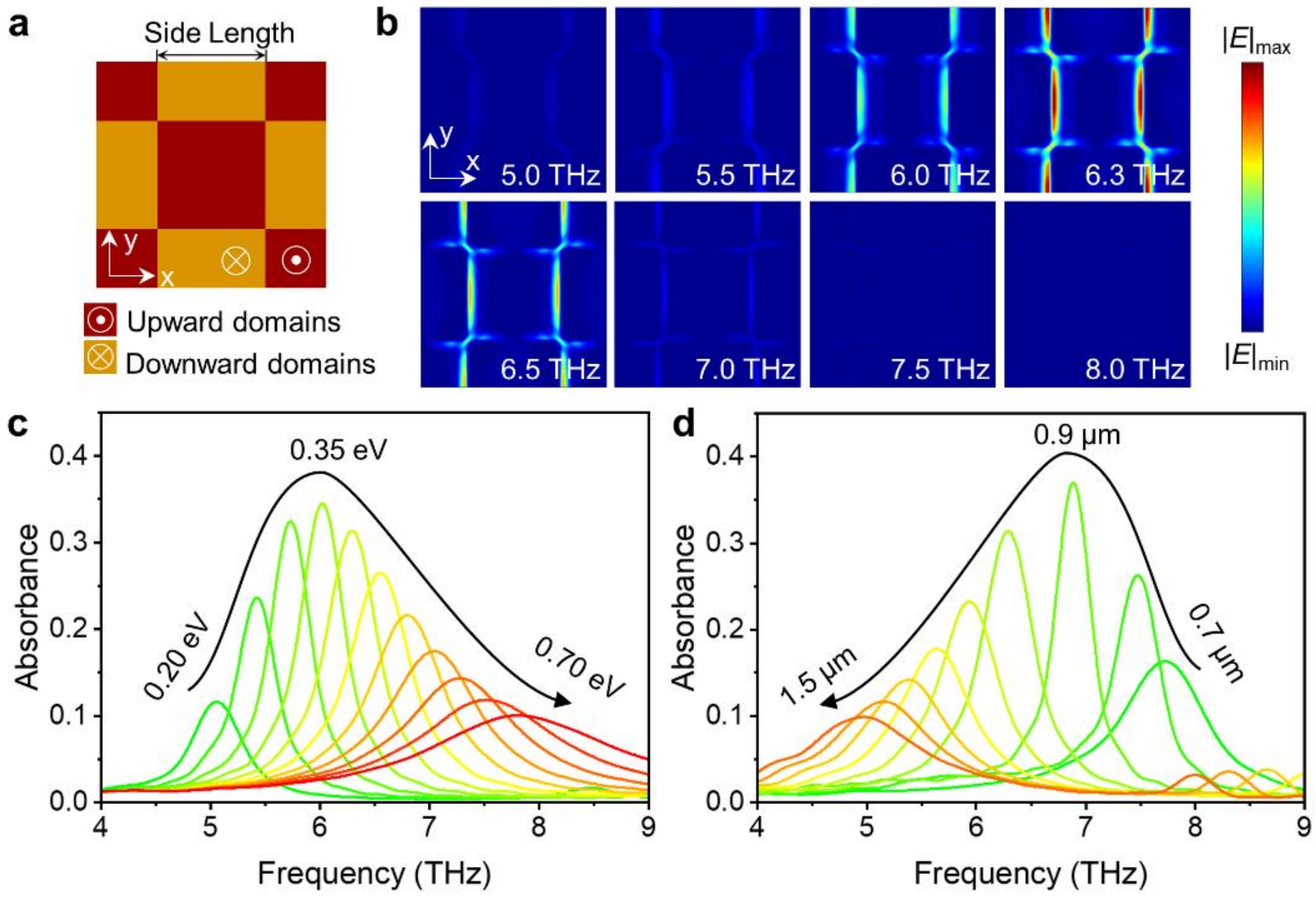

**Fig. 2.** Optical properties of graphene-based terahertz detector. (a) The schematic of the shape of a simulation unit. (b) Intensity of electric field at graphene layer under different frequency of incident wave. Absorption under the incident terahertz wave whose frequency ranges from 4 to 9 THz depends on (c) the chemical potential of graphene and (d) the side length of the ferroelectric domains.

Furthermore, the electrical property of graphene-based terahertz detector is investigated as shown in Fig. 3. Besides the optical conductivity of graphene influenced by incident waves and the interaction between graphene and polarized ferroelectric domains, the size of graphene is a vital factor to determine the electrical property of proposed device. To meet the requirement of periodic boundary condition and scale the area of graphene down as much as possible, the whole graphene is designed as square shape with 60 μm side length of the whole graphene.

Fig. 3(a) shows the *I-V* curve with and without 6.30 THz incident waves whose power is 1 μW by applying a bias voltage ranging from 0 to 1 V. In detail, since the number of photo-generated carriers is definite under 1 μW 6.30 THz waves contributing to illuminated current, the difference between illuminated current and dark current will reach saturation as the voltage increasing. As a result, it features that the illuminated *I-V* curve is not linear near 0.2 V. With bias voltage ranging from −0.15 to 0.15 V, both dark and illuminated currents are linear, which shows the photoconductive behavior near zero bias voltage and a stable on-off ratio (6.6 at 0.1 V) of the device. This behavior is attributed to carriers tunneling at the intersection of four adjacent ferroelectric domains as shown in Fig. 1(b). This explanation can be proved by the ultrahigh hole current density near the intersection of two adjacent downward domains in Fig. 3(b). Also, the electron current density has similar distribution on upward domains.

Fig. 3(c) shows the external quantum efficiency (EQE) and normalized detectivity ($D^*$) under the same condition as that in Fig. 3(a). EQE is defined as the ratio of the amounts of electrons collected by electrode to the amounts of incident photons and $D^*$ is defined as the signal to noise ratio in the condition of per unit surface area, bandwidth and incident power. Here, we calculate EQE and $D^*$ according to the responsivity, as $\mathrm{EQE}=Rh\nu/e$ and $D^*=RS^{1/2}/(eI_D)^{1/2}$, where $\nu$ is the frequency of incident wave, $S$ is the surface area and $h$ is the Planck constant. The responsivity ($R$) which is defined as the efficiency of a photodetector converting the incident light into electric current, follows the formula as $R=(I_L-I_D)/P_{in}$, where $I_L$ is the photocurrent, $I_D$ is the dark current and $P_{in}$ is the power of incident wave. It can be distinctly realized that EQE has the same tendency as responsivity according to the equation and the decrease of $D^*$ after 0.42 V bias voltage is due to linear growth of dark current and the saturation of the difference of illuminated current and dark current. The maximum EQE is 51.83% at 1 V bias and the maximum $D^*$ at 0.42 V bias voltage is $1.16\times10^{11}$ Jones.

As shown in Fig. 3(d), the responsivity is remarkably enhanced without shift of absorption peak as bias voltage increasing. The maximum responsivity at 1 V bias voltage is 19.89 A $W^{-1}$. We consider about the performance of the device and optimize the working voltage as 0.6 V. When the power of 6.30 THz incident wave is 1 μW and the bias voltage is 0.6 V, the responsivity, on-off ratio, EQE and $D^*$ of proposed device are 17.56 A $W^{-1}$, 4.13, 45.76%, $1.11\times10^{11}$ Jones, respectively. This type of device has potential to get lower dark current and faster response with the combination to other two-dimensional materials forming a heterostructure. Additionally, we have summarized some graphene THz detectors with advanced performance as shown in Table 1 and calculate NEP for comparison following the formula as NEP= $S^{1/2}/D^*$.

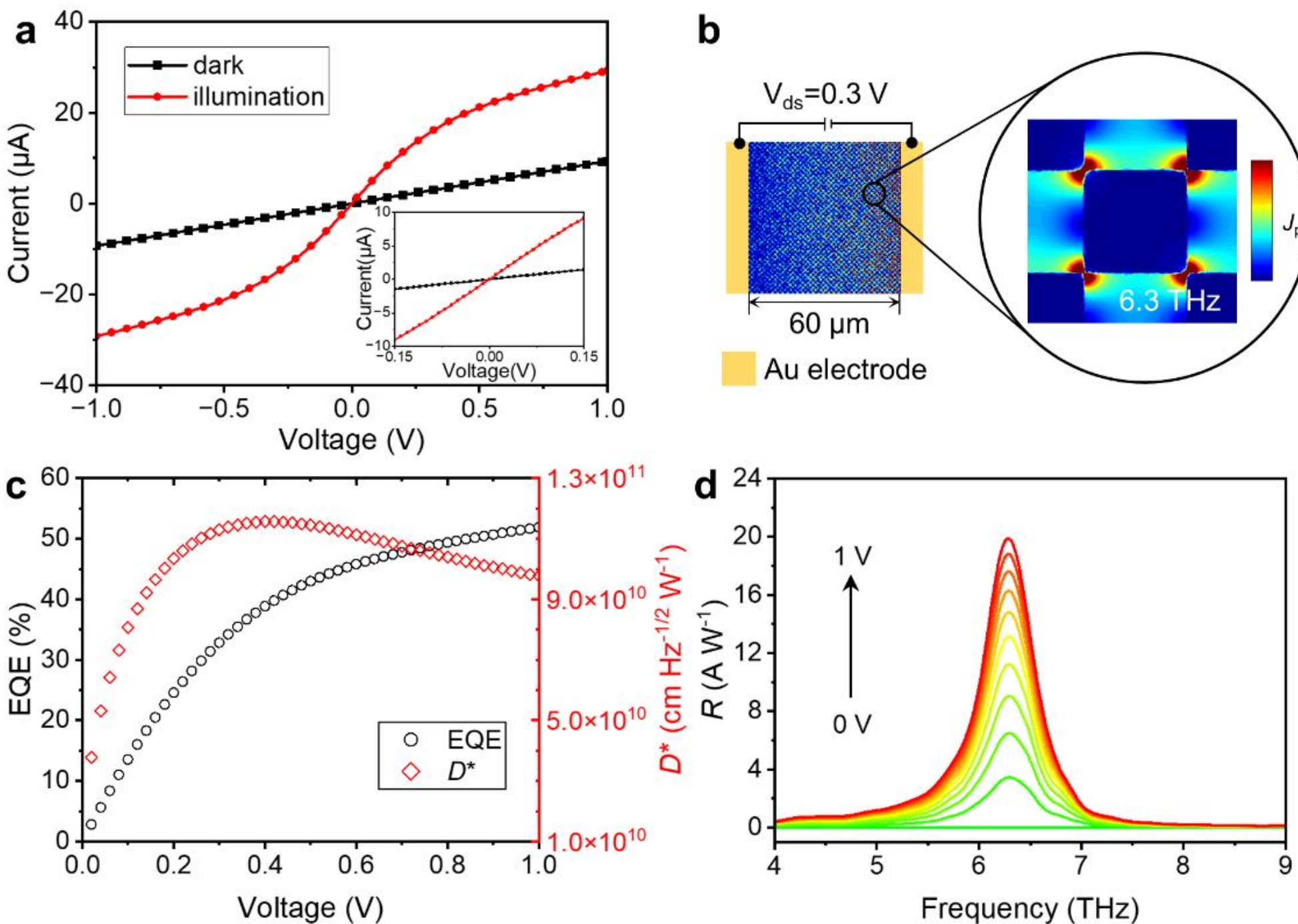

**Fig. 3.** Electrical properties of graphene-based terahertz detector. (a) The dark current and photo current of the device under −1 to 1 V bias voltage. (b) The left figure is hole current density distribution of graphene at 0.6 V under 6.30 THz wave. The right one is the detail in one unit. (c) The EQE and the normalized Detectivity ($D^*$) as the function of the bias voltage under 6.30 THz incident wave. (d) The Responsivity ($R$) as the function of frequency and the bias voltage ranging from 0 to 1 V.

**Table. 1.** Performance comparison of graphene-based THz detectors

| Operating Frequency (THz) | Responsivity | NEP (pW $Hz^{-1/2}$) | Reference |
|---|---|---|---|
| 0.075-0.11 | 976 V $W^{-1}$ | 2.87 | [39] |
| 0.3, 2.5-4.7 | 0.29 A $W^{-1}$, 1.8 mA $W^{-1}$ | ~ | [14] |
| 0.13 | 20 V $W^{-1}$ | $6\times10^5$ | [40] |
| 0.11 | 22.46 mA $W^{-1}$ | 64 | [41] |
| 1.8-4.25 | 25 mA $W^{-1}$ (105V $W^{-1}$) | 80 | [15] |
| 2.5 | 10 V $W^{-1}$ | 1100 | [42] |
| 4.97-7.85 | 17.56 A $W^{-1}$ | 54.05 | This work |

Benefiting from the non-volatile polarization, we can add a small back-gate voltage on the device while maintaining the predetermined polarization of the ferroelectric layer [26, 43]. Considering structure of graphene-BFO-electrode as a perfect capacitor and Eqn. (2), here we can find the variation of the chemical potential of graphene influenced by adding a small back gate ($V_{bg}$) as shown in Eqn. (6)[16].

$$\mu_c = \hbar v_f \left( \pi N_0 + \pi C_d V_{bg} \right)^{1/2} \tag{6}$$

As shown in Fig. 4(a) and Fig. 4(b), it shows a tunable absorption and responsivity from 6.23 to 6.50 THz by adding an additional back-gate voltage and the minimum shift of absorption peak is 0.02 THz when the variation of chemical potential (Δ) changes from 0.02 to 0.03 eV. These features reveal the great potential in portable spectrometer combining the mathematical algorithms for spectrum reconstruction whose resolution depends on the gradient of back-gate voltage. Here is the formula express the measured photocurrents ($I_n$) of devices with different SL, $\mu_c$ and $V_{bg}$ under a specific unknown spectrum ($S$) [44, 45],

$$I_n = \int_{\lambda} S(\delta) R_n(\delta) \mathrm{d}\delta \tag{7}$$

where $\lambda$ presents the operating wavelength range, $R_n$ is the responsivity of the device in different conditions. As shown in Fig. 4(c), according to the responsivities and measured photocurrent under various conditions in one device, the spectrum can be finally reconstructed with the resolution higher than 0.02 THz and a wide wave band of 4.97 to 7.85 THz.

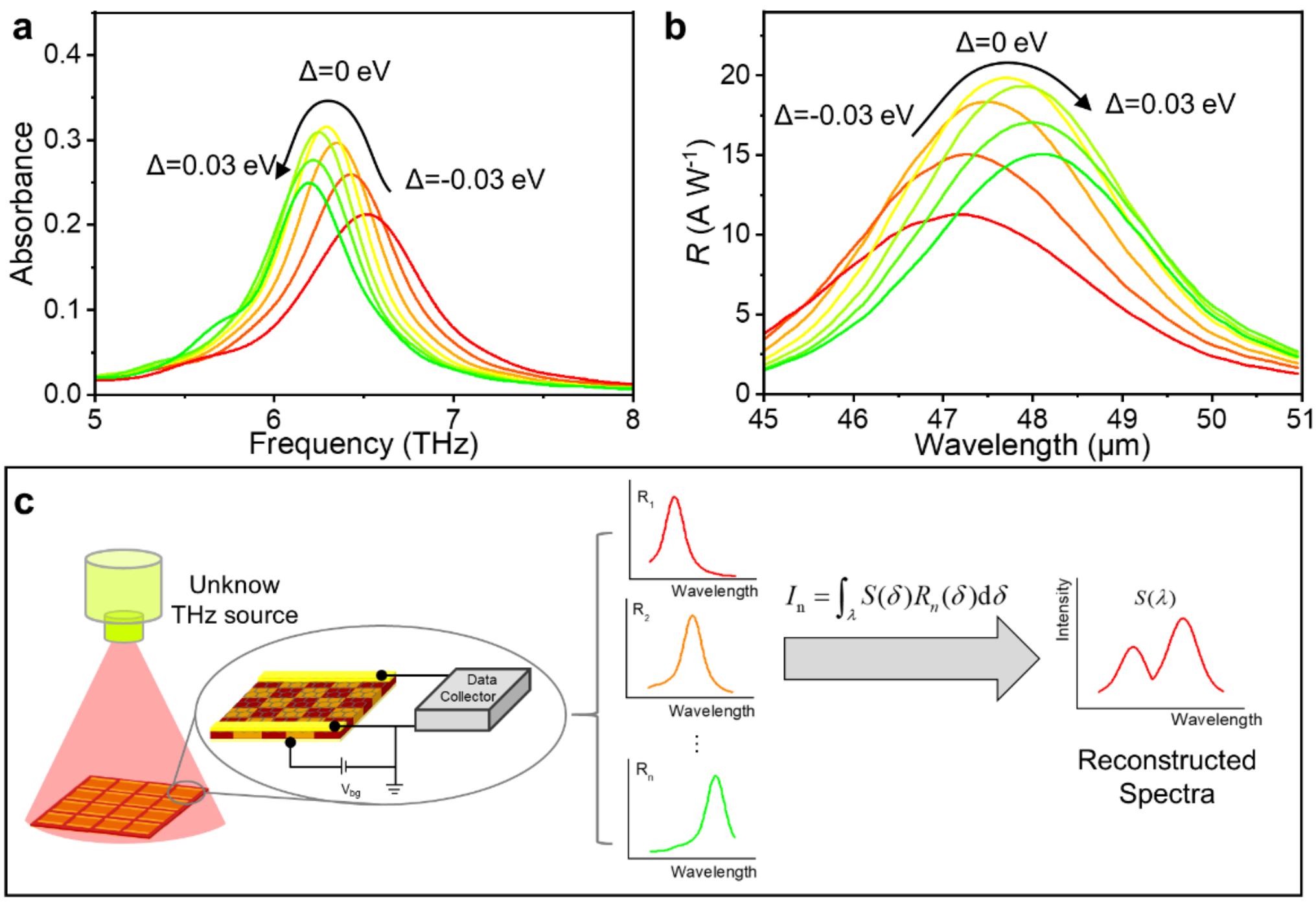

**Fig. 4.** Promising application for portable spectrometer. The (a) absorbance and (b) responsivity of the device tuning by a small back-gate voltage, where Δ is the variation of $\mu_c$. (c) Schematic of spectrum reconstruction based on ferroelectric-integrated graphene plasmonic THz detectors array with wide spectral responses.

In summary, we demonstrate an adaptive terahertz detector by integrating single-layer graphene with periodically polarized ferroelectric thin film. It shows a strong SPR effect excited by periodic polarized ferroelectric domains array, resulting an ultra-high responsivity up to 17.56 A $W^{-1}$ and normalized detectivity up to $1.11\times10^{11}$ Jones in terahertz frequencies. Importantly, our device owns an adaptive resonant absorption peak ranging from 4.97 to 7.85 THz by reconfiguring the periodic polarized ferroelectric domains array and precisely controlled by a tiny gate voltage. These features show the excellent physical properties and fascinating potentials of highly efficient photo-induced-carriers generation. In addition, our device is appropriate for portable spectrometer applications with reconstruct spectrum from 4.97 to 7.85 THz and resolution higher than 0.02 THz. This novel conception of ferroelectric-integrated graphene THz detector with high sensitivity, broadband response and adaptive absorption, as well as easy integration, offers a promising route for smart-production of new optoelectronic devices and potentially pave the practical applications for THz portable spectrometer.

## Competing interests

The authors declare no competing interests.

## Data availability

The data that support the findings of this study are available from the corresponding author upon reasonable request.

## Acknowledgements

This work was financially supported by the National Natural Science Foundation of China (Grant Nos. 62371095 and 62201096), the National Key Research and Development Program of China (Grant No. 2022YFB3206100), the Key R&D Program of Sichuan Province (Grant Nos. 2022JDTD0020, and 2022YFG0163), the Natural Science Foundation of Sichuan Province (Grant No. 2024NSFSC0509), and the China Postdoctoral Science Foundation (Grant No. 2024T170097).

## References

[1] T. Araki, K. Li, D. Suzuki, T. Abe, R. Kawabata, T. Uemura, S. Izumi, S. Tsuruta, N. Terasaki, Y. Kawano, T. Sekitani, Broadband Photodetectors and Imagers in Stretchable Electronics Packaging. Advanced materials (Deerfield Beach, Fla.) e2304048 (2023)

[2] H. Sarieddeen, M. S. Alouini, T. Y. Al-Naffouri, An Overview of Signal Processing Techniques for Terahertz Communications. Proceedings of the Ieee 109, 1628-1665 (2021)

[3] M. S. Shur, Terahertz Plasmonic Technology. Ieee Sensors Journal 21, 12752-12763 (2021)

[4] H. G. Wang, F. F. Zheng, Y. H. Xu, M. G. Mauk, X. B. Qiu, Z. Tian, L. L. Zhang, Recent progress in terahertz biosensors based on artificial electromagnetic subwavelength structure. Trac-Trends in Analytical Chemistry 158, (2023)

[5] Z. Y. Yan, L. G. Zhu, K. Meng, W. X. Huang, Q. W. Shi, THz medical imaging: from in vitro to in vivo. Trends in Biotechnology 40, 816-830 (2022)

[6] J. Guo, S. Gu, L. Lin, Y. Liu, J. Cai, H. Cai, Y. Tian, Y. Zhang, Q. Zhang, Z. Liu, Y. Zhang, X. Zhang, Y. Lin, W. Huang, L. Gu, J. Zhang, Type-printable photodetector arrays for multichannel meta-infrared imaging. Nature Communications 15, (2024)

[7] J. X. Guo, S. D. Li, Z. B. He, Y. Y. Li, Z. C. Lei, Y. Liu, W. Huang, T. X. Gong, Q. Q. Ai, L. N. Mao, Y. W. He, Y. Z. Ke, S. F. Zhou, B. Yu, Near-infrared photodetector based on few-layer MoS2 with sensitivity enhanced by localized surface plasmon resonance. Applied Surface Science 483, 1037-1043 (2019)

[8] Y. Liu, W. Huang, W. J. Chen, X. W. Wang, J. X. Guo, H. Tian, H. N. Zhang, Y. T. Wang, B. Yu, T. L. Ren, J. Xu, Plasmon resonance enhanced WS2 photodetector with ultra-high sensitivity and stability. Applied Surface Science 481, 1127-1132 (2019)

[9] M. S. Islam, J. Sultana, M. Biabanifard, Z. Vafapour, M. J. Nine, A. Dinovitser, C. M. B. Cordeiro, B. W. H. Ng, D. Abbott, Tunable localized surface plasmon graphene metasurface for multiband superabsorption and terahertz sensing. Carbon 158, 559-567 (2020)

[10] J. Chen, Y. Liu, S. Li, L. Lin, Y. Li, W. Huang, J. Guo, Ferroelectric-controlled graphene plasmonic surfaces for all-optical neuromorphic vision. Science China Technological Sciences 67, 765-773 (2024)

[11] J. Guo, L. Lin, S. Li, J. Chen, S. Wang, W. Wu, J. Cai, T. Zhou, Y. Liu, W. Huang, Ferroelectric superdomain controlled graphene plasmon for tunable mid-infrared photodetector with dual-band spectral selectivity. Carbon 189, 596-603 (2022)

[12] J. A. Delgado-Notario, W. Knap, V. Clerico, J. Salvador-Sanchez, J. Calvo-Gallego, T. Taniguchi, K. Watanabe, T. Otsuji, V. V. Popov, D. V. Fateev, E. Diez, J. E. Velazquez-Perez, Y. M. Meziani, Enhanced terahertz detection of multigate graphene nanostructures. Nanophotonics 11, 519-529 (2022)

[13] B. C. Yao, Y. Liu, S. W. Huang, C. Choi, Z. D. Xie, J. F. Flores, Y. Wu, M. B. Yu, D. L. Kwong, Y. Huang, Y. J. Rao, X. F. Duan, C. W. Wong, Broadband gate-tunable terahertz plasmons in graphene heterostructures. Nature Photonics 12, 22-+ (2018)

[14] J. M. Caridad, Ó. Castelló, S. M. López Baptista, T. Taniguchi, K. Watanabe, H. G. Roskos, J. A. Delgado-Notario, Room-Temperature Plasmon-Assisted Resonant THz Detection in Single-Layer Graphene Transistors. Nano Letters 24, 935-942 (2024)

[15] S. Castilla, B. Terrés, M. Autore, L. Viti, J. Li, A. Y. Nikitin, I. Vangelidis, K. Watanabe, T. Taniguchi, E. Lidorikis, M. S. Vitiello, R. Hillenbrand, K.-J. Tielrooij, F. H. L. Koppens, Fast and Sensitive Terahertz Detection Using an Antenna-Integrated

Graphene pn Junction. Nano Letters 19, 2765-2773 (2019)

[16] B. Liu, Y. Peng, Y. Hao, Y. Zhu, S. Chang, S. Zhuang, Ultra-wideband terahertz fingerprint enhancement sensing and inversion model supported by single-pixel reconfigurable graphene metasurface. PhotoniX 5, (2024)

[17] Q. S. Guo, R. W. Yu, C. Li, S. F. Yuan, B. C. Deng, F. J. G. de Abajo, F. N. Xia, Efficient electrical detection of mid-infrared graphene plasmons at room temperature. Nature Materials 17, 986-+ (2018)

[18] A. Rajapitamahuni, J. Hoffman, C. H. Ahn, X. Hong, Examining Graphene Field Effect Sensors for Ferroelectric Thin Film Studies. Nano Letters 13, 4374-4379 (2013)

[19] M. V. Strikha, A. N. Morozovska, Limits for the graphene on ferroelectric domain wall p-n-junction rectifier for different regimes of current. Journal of Applied Physics 120, (2016)

[20] T. E. Beechem, M. D. Goldflam, M. B. Sinclair, D. W. Peters, A. E. McDonald, E. A. Paisley, A. R. Kitahara, D. E. Drury, D. B. Burckel, P. S. Finnegan, J. W. Kim, Y. Choi, P. J. Ryan, J. F. Ihlefeld, Tunable Infrared Devices via Ferroelectric Domain Reconfiguration. Advanced Optical Materials 6, (2018)

[21] S. H. Baek, H. W. Jang, C. M. Folkman, Y. L. Li, B. Winchester, J. X. Zhang, Q. He, Y. H. Chu, C. T. Nelson, M. S. Rzchowski, X. Q. Pan, R. Ramesh, L. Q. Chen, C. B. Eom, Ferroelastic switching for nanoscale non-volatile magnetoelectric devices. Nature Materials 9, 309-314 (2010)

[22] H. Lu, C. W. Bark, D. E. de los Ojos, J. Alcala, C. B. Eom, G. Catalan, A. Gruverman, Mechanical Writing of Ferroelectric Polarization. Science 336, 59-61 (2012)

[23] Y. Tian, L. Y. Wei, Q. H. Zhang, H. B. Huang, Y. L. Zhang, H. Zhou, F. J. Ma, L. Gu, S. Meng, L. Q. Chen, C. W. Nan, J. X. Zhang, Water printing of ferroelectric polarization. Nature Communications 9, (2018)

[24] K. S. Novoselov, A. K. Geim, S. V. Morozov, D. Jiang, Y. Zhang, S. V. Dubonos, I. V. Grigorieva, A. A. Firsov, Electric field effect in atomically thin carbon films. Science 306, 666-669 (2004)

[25] Y. Si, E. T. Samulski, Synthesis of water soluble graphene. Nano Letters 8, 1679-1682 (2008)

[26] V. Nagarajan, A. Roytburd, A. Stanishevsky, S. Prasertchoung, T. Zhao, L. Chen, J. Melngailis, O. Auciello, R. Ramesh, Dynamics of ferroelastic domains in ferroelectric thin films. Nature Materials 2, 43-47 (2003)

[27] J. X. Guo, Y. Liu, Y. Lin, Y. Tian, J. X. Zhang, T. X. Gong, T. D. Cheng, W. Huang, X. S. Zhang, Simulation of tuning graphene plasmonic behaviors by ferroelectric domains for self-driven infrared photodetector applications. Nanoscale

11, 20868-20875 (2019)

[28] A. Vakil, N. Engheta, Transformation Optics Using Graphene. Science 332, 1291-1294 (2011)

[29] K. F. Mak, L. Ju, F. Wang, T. F. Heinz, Optical spectroscopy of graphene: From the far infrared to the ultraviolet. Solid State Communications 152, 1341-1349 (2012)

[30] L. A. Falkovsky, S. S. Pershoguba, Optical far-infrared properties of a graphene monolayer and multilayer. Physical Review B 76, (2007)

[31] L. A. Falkovsky In *Optical properties of graphene*, International Conference on Theoretical Physics (Dubna-Nano2008), JINR, Bogoliubov Lab Theoret Phys, Dubna, RUSSIA, Jul 07-11, 2008; JINR, Bogoliubov Lab Theoret Phys, Dubna, RUSSIA, 2008.

[32] L. A. Falkovsky, A. A. Varlamov, Space-time dispersion of graphene conductivity. European Physical Journal B 56, 281-284 (2007)

[33] G. W. Hanson, Dyadic Green’s functions and guided surface waves for a surface conductivity model of graphene. Journal of Applied Physics 103, 064302 (2008)

[34] Z. Yi, C. Liang, X. Chen, Z. Zhou, Y. Tang, X. Ye, Y. Yi, J. Wang, P. Wu, Dual-Band Plasmonic Perfect Absorber Based on Graphene Metamaterials for Refractive Index Sensing Application. Micromachines 10, 443 (2019)

[35] Q. L. Bao, K. P. Loh, Graphene Photonics, Plasmonics, and Broadband Optoelectronic Devices. Acs Nano 6, 3677-3694 (2012)

[36] S. A. Mikhailov, K. Ziegler, New electromagnetic mode in graphene. Physical Review Letters 99, (2007)

[37] V. P. Gusynin, S. G. Sharapov, J. P. Carbotte, Magneto-optical conductivity in graphene. Journal of Physics-Condensed Matter 19, (2007)

[38] A. I. Kurchak, E. A. Eliseev, S. V. Kalinin, M. V. Strikha, A. N. Morozovska, p-n Junction Dynamics Induced in a Graphene Channel by Ferroelectric-Domain Motion in the Substrate. Physical Review Applied 8, (2017)

[39] Y. Yao, M. Cai, J. Fu, S. Hou, Y. Cai, F. He, X. Guo, Y. Zhu, Configurable microcavity-enhanced graphene photothermoelectric terahertz detectors. Photonics Research 12, 2300-2310 (2024)

[40] D. A. Bandurin, I. Gayduchenko, Y. Cao, M. Moskotin, A. Principi, I. V. Grigorieva, G. Goltsman, G. Fedorov, D. Svintsov, Dual origin of room temperature sub-terahertz photoresponse in graphene field effect transistors. Applied Physics Letters 112, 141101 (2018)

[41] Y. He, N. Fu, M. Jiang, X. Lv, S. Guo, L. Han, L. Zhang, B. Zhao, G. Chen, X. Chen, L. Wang, Carbon-based self-powered terahertz detector with multi-walled carbon nanotubes-graphene heterostructure. Carbon 221, 118886 (2024)

[42] X. Cai, A. B. Sushkov, R. J. Suess, M. M. Jadidi, G. S. Jenkins, L. O. Nyakiti, R. L. Myers-Ward, S. Li, J. Yan, D. K. Gaskill, T. E. Murphy, H. D. Drew, M. S. Fuhrer, Sensitive room-temperature terahertz detection via the photothermoelectric effect in graphene. Nature Nanotechnology 9, 814-819 (2014)

[43] L. Lv, F. W. Zhuge, F. J. Xie, X. J. Xiong, Q. F. Zhang, N. Zhang, Y. Huang, T. Y. Zhai, Reconfigurable two-dimensional optoelectronic devices enabled by local ferroelectric polarization. Nature Communications 10, (2019)

[44] H. H. Yoon, H. A. Fernandez, F. Nigmatulin, W. W. Cai, Z. Y. Vane, H. X. Cui, F. Ahmed, X. Q. Cui, M. G. Uddin, E. D. Minot, H. Lipsanen, K. Kim, P. Hakonen, T. Hasan, Z. P. Sun, Miniaturized spectrometers with a tunable van der Waals junction. Science 378, 396-399 (2022)

[45] J. R. Wen, L. Y. Hao, C. Gao, H. L. Wang, K. Mo, W. J. Yuan, X. Chen, Y. S. Wang, Y. G. Zhang, Y. C. Shao, C. Y. Yang, W. D. Shen, Deep Learning-Based Miniaturized All-Dielectric Ultracompact Film Spectrometer. Acs Photonics (2022)